\documentstyle[aps,twocolumn,epsfig]{revtex}
\begin{document}
\twocolumn[\hsize\textwidth\columnwidth\hsize\csname
@twocolumnfalse\endcsname
\draft

\title{Ferromagnetism in Insulating, Doped Diluted Magnetic
Semiconductors}

\author{Xin Wan and R. N. Bhatt}
\address{Department of Electrical Engineering, Princeton University,
        Princeton, NJ 08544-5263}
\date{\today}
\maketitle
\begin{abstract}

We report results of a Monte Carlo study of doped, diluted magnetic 
semiconductors in the low carrier density (insulating) regime. We
find that the system undergoes a transition from a paramagnet
at high temperatures to a ferromagnet at low temperatures.
However, in strong contrast to uniform systems, disorder effects
dominate the entire collective behavior, leading to very unconventional
properties such as (1) magnetization curves that cannot be described by
theories
based on critical fluctuations, or on spin wave analysis
over any significant fraction of the phase diagram; (2) a large
peak in susceptibilty well below the ordering temperature;
and (3) specific heat curves that point out the inadequacy of
a classical Heisenberg model for spin-5/2 Mn ions. A picture of
percolating magnetic polarons appears to describe the
data well, and leads to a prescription for correcting the
unphysical results of the classical continuous spin model in
terms of a discrete vector model.\\ 

\end{abstract}

]

In diluted magnetic semiconductors (DMS), such as Zn$_{1-x}$Mn$_x$Te and
Cd$_{1-x}$Mn$_x$Se, carriers localized at impurities can induce sizable
magnetic moments on the scale of their hydrogen-like orbits. 
These entities, known as bound magnetic polarons (BMPs), are
created by the $sp$-$d$ exchange interactions of localized carriers with
magnetic ions in their vicinity.  
The mechanism and thermodynamics of mutually independent BMPs in lightly
doped DMS are well understood~\cite{review,averous}. 
However, experiments on doped, p-type ZnMnTe~\cite{liu} at intermediate
doping
densities (but clearly below the transition to a metallic state), have
shown a
significant enhancement in susceptibility at low temperatures ($T$),
implying a weak ferromagnetic polaron-polaron interaction. 
Based on an idealized, but exactly soluable, model of the single
polaron, Wolff {\it et
al.}~\cite{wolff} concluded that Mn$^{2+}$ spins in the intervening region
between
polarons formed around two bound carriers can generate an
effective ferromagnetic interaction between the two polarons which
likely
overcomes the direct antiferromagnetic interactions between impurities 
(due to virtual carrier hopping, as in usual nonmagnetic
semiconductors).
 
In this paper, we report a Monte Carlo study of a more
realistic model (with spatially dependent carrier wavefunctions) 
and demonstrate the generation of such effective
ferromagnetic interactions within a model with only
antiferromagnetic exchange interactions. We find that the system
undergoes a transition to a ferromagnetic state below $T=T_c$ ; however,
the
state is characterized by a very inhomogeneous spatial magnetization
pattern,  
with a large (but finite) peak in the magnetic susceptibility at a
temperature $T_p$
{\it well} (one or two decades) below $T_c$, 
and $ T_p/T_c $ can be tuned, depending on the carrier density.
Furthermore,
the classical (continuous spin) Heisenberg model is found to yield
unphysical
results for the low-T specific heat, which is corrected by
using a discrete vector spin model. 

The $sp$-$d$ exchange interaction of a localized carrier with a magnetic
ion on a distance ${\bf R}$ can be described by the spin Hamiltonian: 
\begin{equation}
        {\cal H} = J_0 |\phi({\bf R})|^2 {\bf s} \cdot {\bf S},
\end{equation}
where $J_0$ is the exchange constant, while ${\bf s}$ and ${\bf S}$ are
spins of the carrier and the magnetic ion, respectively. For this study,
we consider for concreteness the case of electron doping (${\bf s}
= 1/2$); our qualitative results are, however,
valid for both electron and hole doping. 
The orbital wavefunction of the localized carrier $\phi(r)$ is taken to
be
of a hydrogenic form $\phi(r) = e^{-r/a_B}$ for simplicity in which
$a_B$ is the Bohr radius of the hydrogen-like orbit; extending to other
spatial forms is straightforward.

On a zincblende lattice, 
Mn ions are introduced randomly to replace cations (one of the two
interpenetrating fcc lattices) while the dopants replace anions randomly
on the other sublattice.  
The ratio of Mn ions to dopants is taken to be 32:1, which corresponds
roughly to 0.1\% Mn at a doping concentration of $10^{18} cm^{-3}$. 
The Bohr radius $a_B$ is taken at a typical value of $20 \AA$ and the
lattice constant $a_0 = 5\AA$.  
Following  Wolff {\it et al.}~\cite{wolff}, the direct 
coupling between carrier spins has been neglected.
In fact, we have found that introducing such a coupling  at the
densities
studied has no qualitative effects on the
thermodynamic phase diagram, and very minor quantitataive changes.  
We also exclude direct Mn-Mn interactions among 0.1\% substitutional
manganese ions,  
since only 1\% of these Mn ions have nearest neighbors. 
(The main effect of these pairs or larger clusters is to
effectively reduce the percentage of active Mn spins, since most cluster
interactions reduce the net magnetic moment - e.g. pairs form inert
singlets). 
The Hamiltonian can be written as 
\begin{equation}
        {\cal H} = \sum_{i,j} J({\bf r}_i,{\bf R}_j) {\bf s}_i \cdot
{\bf S}_j,
\end{equation}
where ${\bf r}_i$ and ${\bf s}_i$ are position and spin of the $i$-th
carrier spin, while ${\bf R}_j$ and ${\bf S}_j$ those of the $j$-th Mn
spin. 
All the spins are treated as classical Heisenberg spins. 
The exchange between a Mn spin and a carrier spin is given by,
\begin{equation}
        J({\bf r}_i, {\bf R}_j) = J_0 e^{- 2 |{\bf r}_i - {\bf R}_j| /
a_B}.
\end{equation}

We carried out simulations on lattices of linear size $L = 40$ to $80$,
which contain $N_c = 8$ to $64$ dopants and $N_d = 256$ to $2048$ Mn
ions.
We averaged up to 3000 samples per data point, depending on $L$
and $T$. $J_0$ is taken as the unit of $T$.
The equilibration of each system, which consists of as many as 2048 Mn
spins, 
is checked by the following technique. 
We simulate two mutually independent samples with identical locations of
particles but different initial spin configurations.
One sample starts from totally random spin configuration, 
while the other from an ordered state, 
where the electron spins and the Mn spins are oriented ferromagnetically
within each kind but opposite to the other kind.
The evolutions of the two samples correspond to the relaxations from
high $T$ and low $T$ limit respectively. 
We expect to obtain the same results, such as magnetization and energy,
in both samples for sufficient long time $t$, when equilibrium is
reached. 

\begin{figure}[ht]
\centerline{
        \epsfig{figure=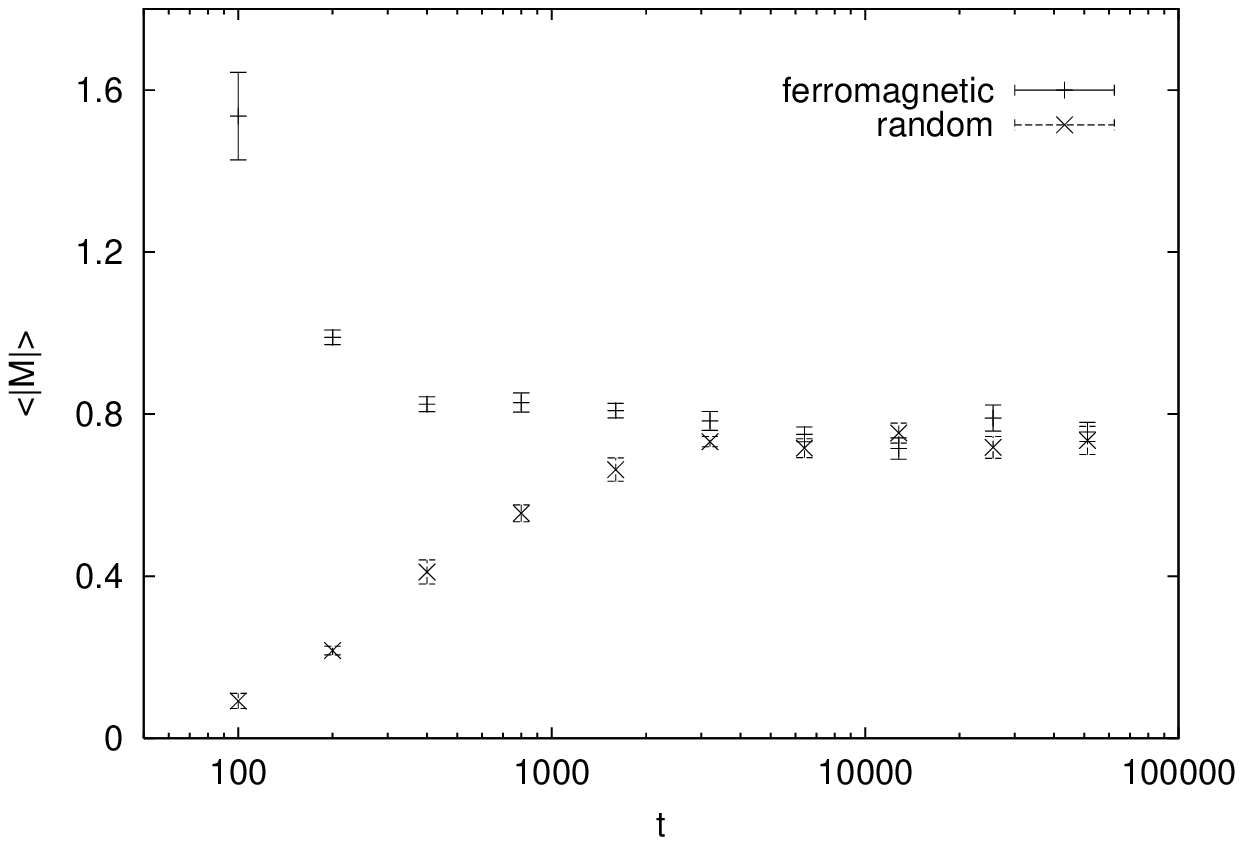,width=3.6in}
}
\caption{
\label{equilibrium}
Plot of magnetization per spin for a system of 256 Mn spins at
$T = 0.01$ as a function of time. The upper curve starts from
ferromagnetic spin configuration, while the lower one from random
configuration. } 
\end{figure}

Figure~\ref{equilibrium} shows the evolution of average magnetization
$\langle |M| \rangle$ with $t$ for 256 Mn spins and temperature $T =
0.01$.
From the convergence of the two curves we can determine the 
equilibration time for that $L$ and $T$.   
Because the magnetization values in the two replicas approach the
equilibrium from different directions,  
we are able to obtain the equilibrium results to desired precision
without wasting precious computer time.  

\begin{figure}[ht]
\centerline{
        \epsfig{figure=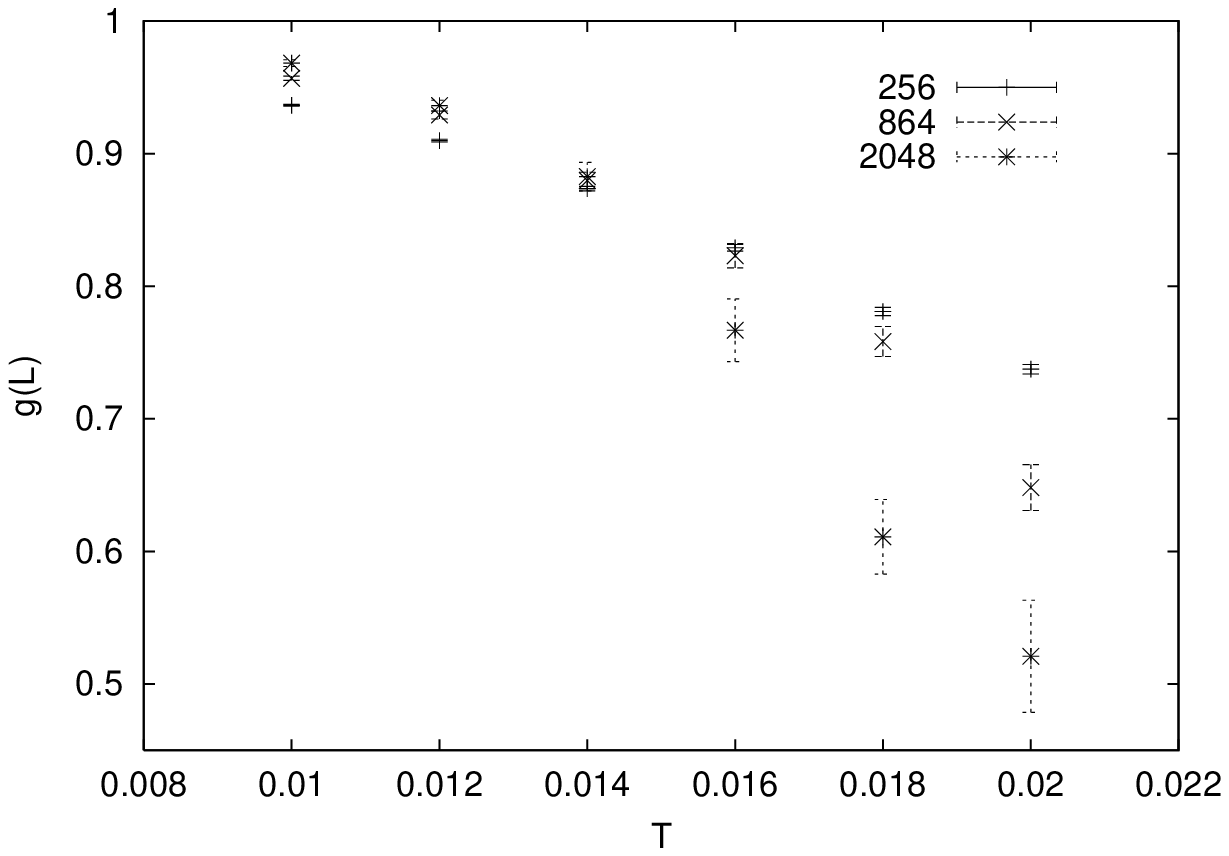,width=3.6in}
}
\caption{ 
\label{cumulant}
Plot of $g(L)$ for system of 256, 864, and 2048 Mn spins at
different temperatures around $T_c = 0.014$. } 
\end{figure}

To determine the critical temperature in the model, we look, in
particular, at the following dimensionless quantity widely used in spin
systems\cite{binder}: 
\begin{equation}
        g(L) = {1 \over 2} \left \{5 - 3 { <|M|^4> \over <|M|^2>^2 }
\right \},
\end{equation}
for system of various sizes (numbers of Mn and electron spins) at
different
temperatures. 
The coefficients in the above equation have been chosen so that $g(L) =
1$ at
$T=0$, and $0$ as $T \rightarrow \infty $. 
In Fig.~\ref{cumulant} we plot $g(L)$ as a function of temperature $T$,
for different number of Mn spins.  
At large $T$, $g(L)$ decreases with number of Mn spins, or system size,
while at low $T$, it increases with system size.
The curves seem to cross around $T = 0.014$, 
at which $g(L)$ is independent of $L$. 
We identify the crossing point as the transition temperature $T_c$, 
as suggested by finite size scaling theory. 

\begin{figure}[ht]
\centerline{
        \epsfig{figure=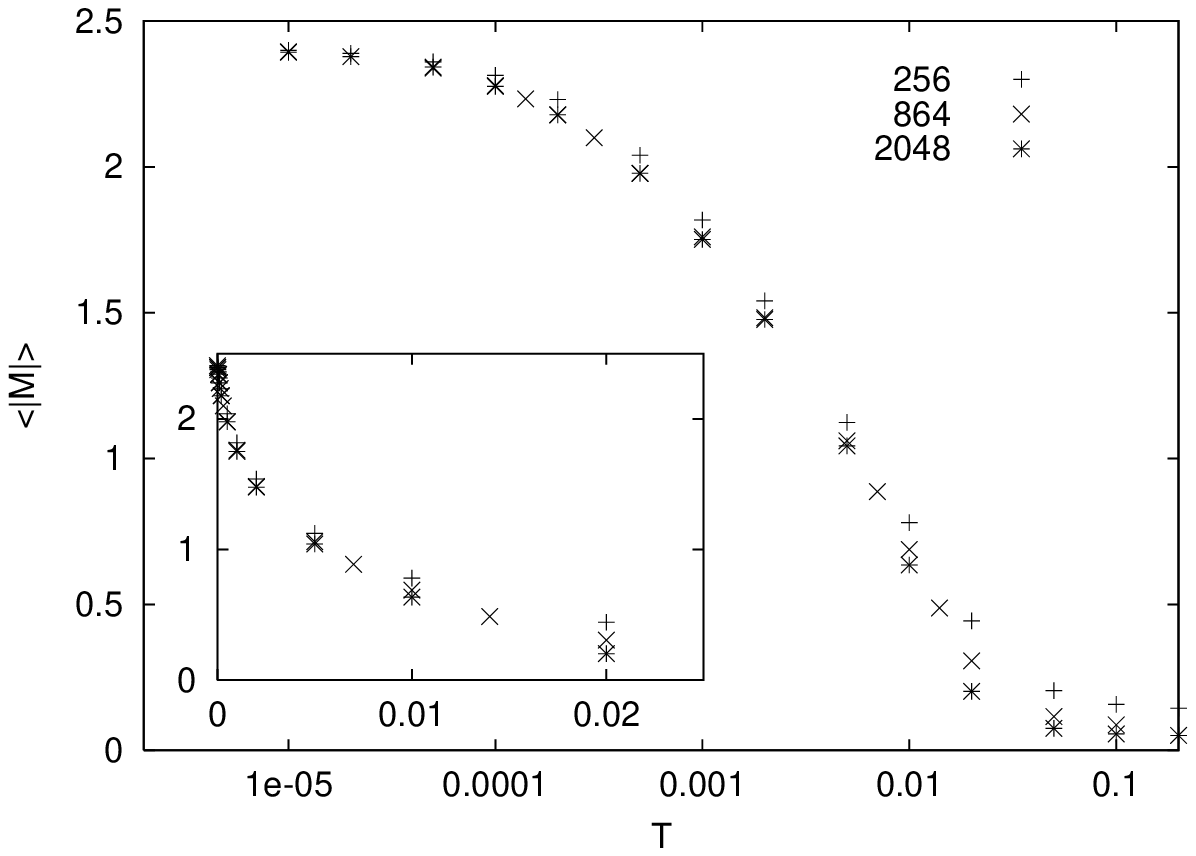,width=3.6in}
}
\caption{ 
\label{magnetization}
Magnetization per spin versus temperature for system of 864
and 2048 Mn spins. } 
\end{figure}

In Fig.~\ref{magnetization} we plot the magnetization per spin at
different temperatures for systems of 864 and 2048 Mn spins. 
Plotted on a logarithmic temperature scale, the magnetization curves
look similar to what we observe in a typical ferromagnet on a linear
plot. 
On a linear plot, however, it is seen that $\langle |M| \rangle$ does
not saturate
even down to $ T_c/5 $. This can be understood as follows. The driving
force
behind the ferromagnetic transition is the ordering of Mn spins around
each localized carrier into magnetic polarons. 
The radius of the polaron surrounding a dopant ion can therefore be
estimated by equalizing the exchange energy at such distance to the
ambient
temperature.  
For localized wavefunctions (in particular of the hydrogenic form),
therefore,
 the radius of a polaron
increases logarithmically with inverse temperature,
$ r_T \sim (a_B/2) \ln (J_0 s S / T) $.
At a certain $T$, the radius of polarons grows large enough so
that polarons percolate into a macroscopic ferromagnetic cluster, whose
formation signals the transition, $ T_c$. 
However, since the percolating cluster does not contain a majority of
the spins until temperatures which are exponentially lower than $ T_c $,
most Mn spins remain free even in the ferromagnetic phase. 
It is these uncoupled spins which give rise to the
slow saturation of magnetization at low $T$, and to the giant
susceptibility
peak below. 
The magnetization appears to approach the saturation value linearly 
at low $T$, within the precision of our simulation. 

\begin{figure}[ht]
\centerline{
        \epsfig{figure=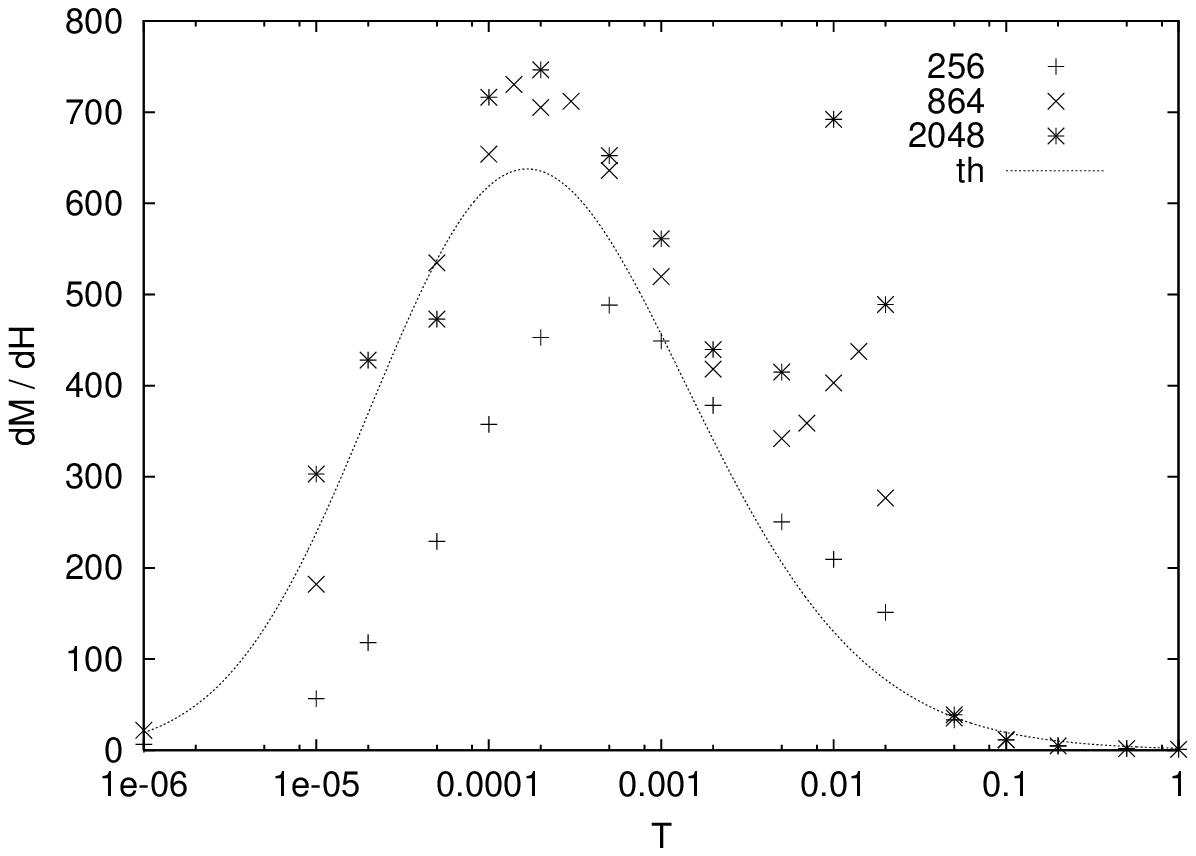,width=3.6in}
}
\caption{ 
\label{susceptibility}
Plot of susceptibility versus temperature $T$ for systems of
various number of Mn spins.  
Solid curve is the estimate of the contributions from manganese spins
outside the percolating polaron cluster. } 
\end{figure}

Figure~\ref{susceptibility} shows the plot of magnetic susceptibility as
a function of $T$ for different system sizes.  
Susceptibility increases dramatically when approaching the critical
temperature $T_c = 0.014$, at which it diverges in the thermodynamic
limit.  
Remarkably, these exists a huge peak at $T = T_p$ two orders of
magnitude below $T_c$, 
which appears to be L-independent for large enough
lattices. 
Further analysis shows the peak at $T_p$ is associated with
single spin excitations of those Mn spins far away from any of the bound
electrons. 

Well below $T_c$, it is convenient to view the
system as a cluster of percolating polarons with a large number of Mn
ions outside 
the cluster, which are interacting weakly with distant electron spins.
Modeling the polarons as rigid shells of radius $r_T$, the number of Mn
ions outside the polaron cluster can be estimated: 
\begin{equation}
        N_T = N_M P(r_T) = N_M e^{-4 \pi \rho_c r_T^3 / 3},
\end{equation}
where $N_M$ is the total number of Mn ions and $\rho_c$ the doping
density. 
$P(r)$ is the probability of a magnetic ion having no bound electrons
within distance $r$.  
Since each of these Mn spins contributes $S^2 / 3 T$ to susceptibility
in low temperature limit, the total susceptibility can be written as: 
\begin{equation}
        \chi_{\rm total} = {N_M S^2 \over 3 T} e^{-4 \pi \rho_c r_T^3 /
3},
\end{equation}
which fits the susceptibility peak in the simulation results reasonably
well as shown in Fig.~\ref{susceptibility}.  
Since finite system size prohibits the existence of manganese spins very
far from any electron spins, the susceptibility peak at lower
temperature
is suppressed in small systems.  
Such an effect can be seen in the smallest system (256 Mn spins) in our
simulation.  

\begin{figure}[ht]
\centerline{
        \epsfig{figure=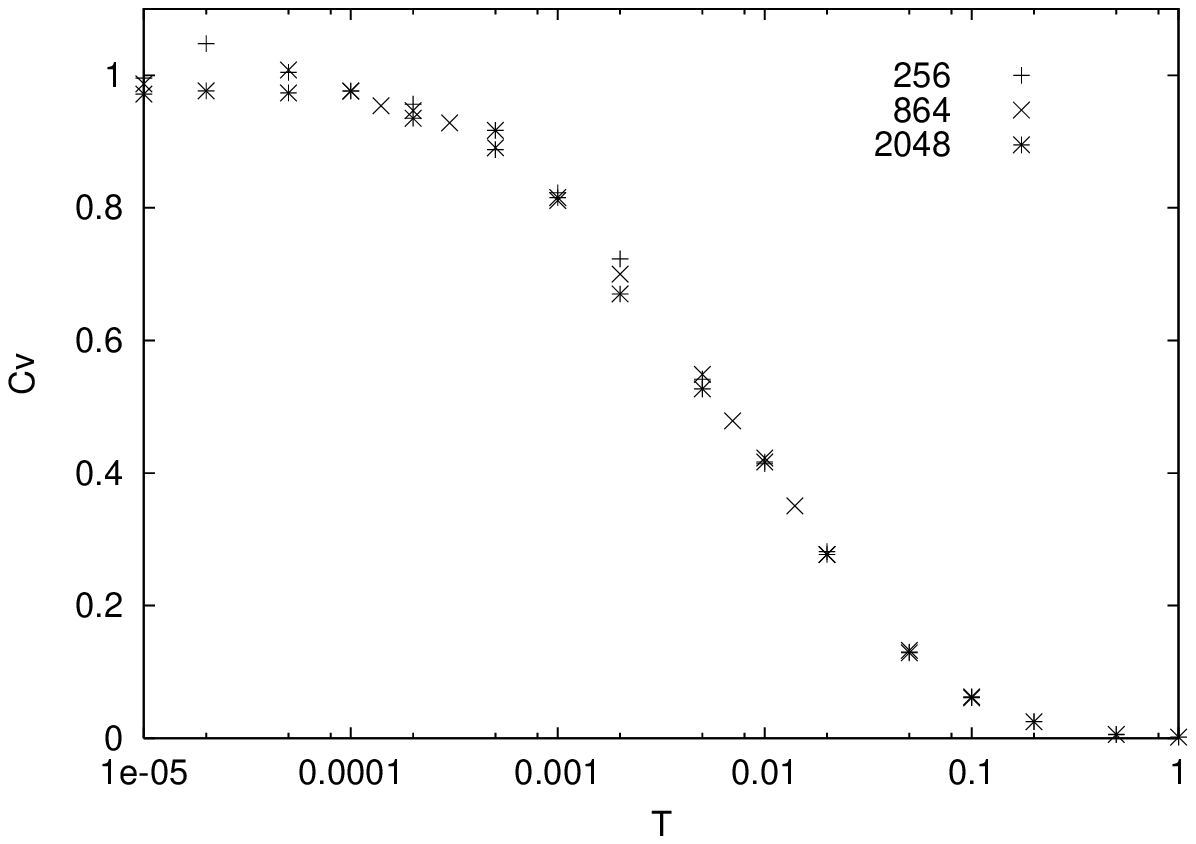,width=3.6in}
}
\caption{ 
\label{cvClassical}
Plot of $Cv$ versus temperature in the classical Heisenberg spin model. }
\end{figure}

In Fig.~\ref{cvClassical}, we show the specific heat per spin as a
function of $T$. 
Notably, the specific heat does not vanish at low $T$,
approaching unity instead. 
This is not surprising for a classical Heisenberg spin model, since the
low
temperature behavior must obey the classical equipartition theorem. 
In order to mimic the quantum mechanical characteristics of Mn spins
without carrying out a full quantum Monte Carlo simulation, we have
adopted a
discrete vector model, in which each Mn spin can be oriented only along
one of the six [100] directions. 
Spin flips in the vector model cost a finite energy therefore are frozen
at very low temperatures.   

Results for a 12-state vector model has been investigated in
more detail and published elsewhere\cite{vertex12}. 
In the two disctrete spin models, the equilibration of a system,
especially
starting with random initial configurations, becomes extremely slow at
low
temperatures with single spin flips, when energy change associated with
such a flip becomes exponentially larger than the temperature. 
In fact, most of the samples are frozen in metastable states.
A cluster algorithm, which is capable of updating large number of spins
(within a polaron or a percolating polaron cluster) in a single step,
has been introduced to speed up the equilibration\cite{vertex12}. 
The algorithm has been found to attain convergence at low
temperatures, especially in small samples. 

\begin{figure}[ht]
\centerline{
        \epsfig{figure=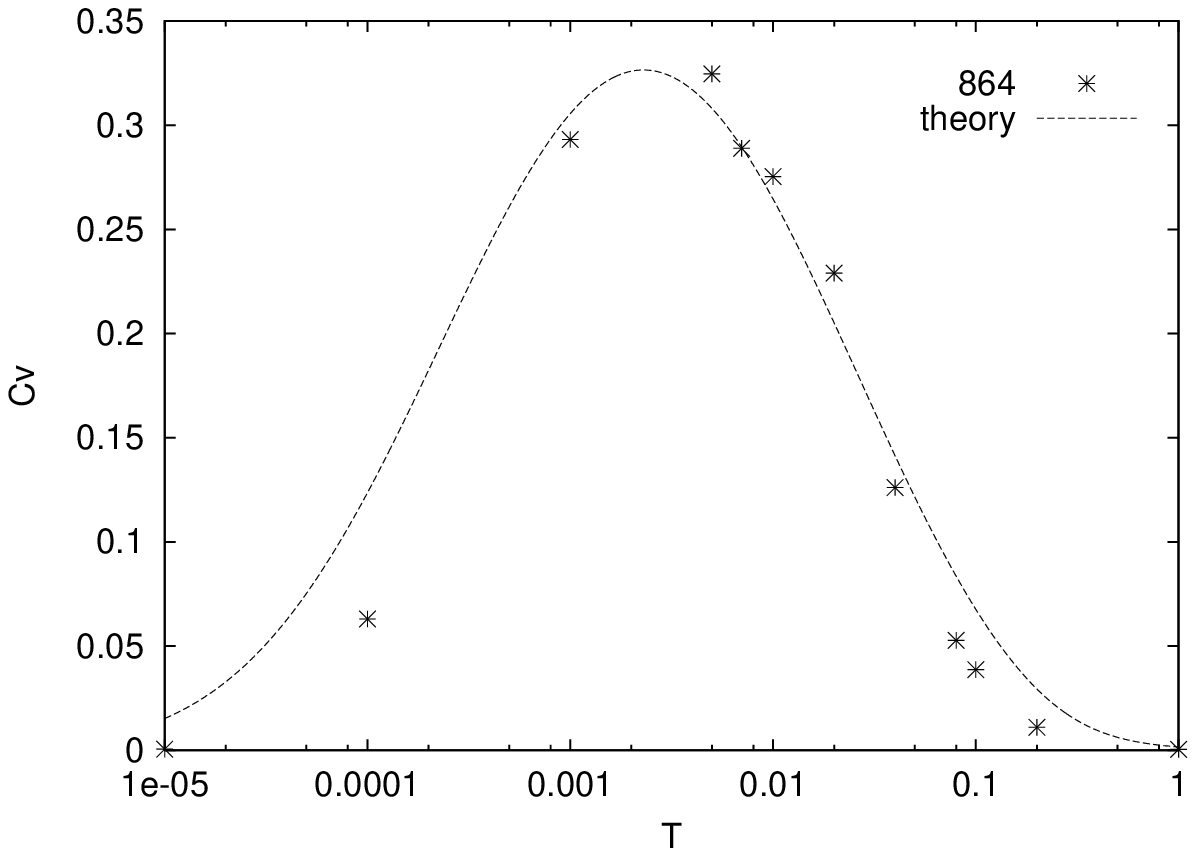,width=3.6in}
}
\caption{ 
\label{cvVertex}
$C_v$ at different temperatures for a system of 864 Mn spins
in the discrete vector model.  
The solid line is the theoretical estimate. }
\end{figure}

The introduction of discrete spin models leads to qualitatively similar
magnetization and susceptibility curves. 
The magnetic susceptibility shows a peak two orders of magnitude
below $T_c$. 
Yet, the discreteness of energy levels simulates the quantization of 
spin energy levels as desired. 
Figure~\ref{cvVertex} shows the plot of specific heat of a system of 864
Mn spins at different temperatures in the 6-vector model.
As a result of discrete symmetry, specific heat per spin drops to zero
in low temperature limit. 
Following a similar argument as for the susceptibility, one can
show~\cite{bhatt}
\begin{equation}
C_v \sim \rho_c a_B r_T^2 e^{-4 \pi \rho_c r_T^3 / 3},
\end{equation}
which fits the simulation results very well. 

\begin{figure}[ht]
\centerline{
        \epsfig{figure=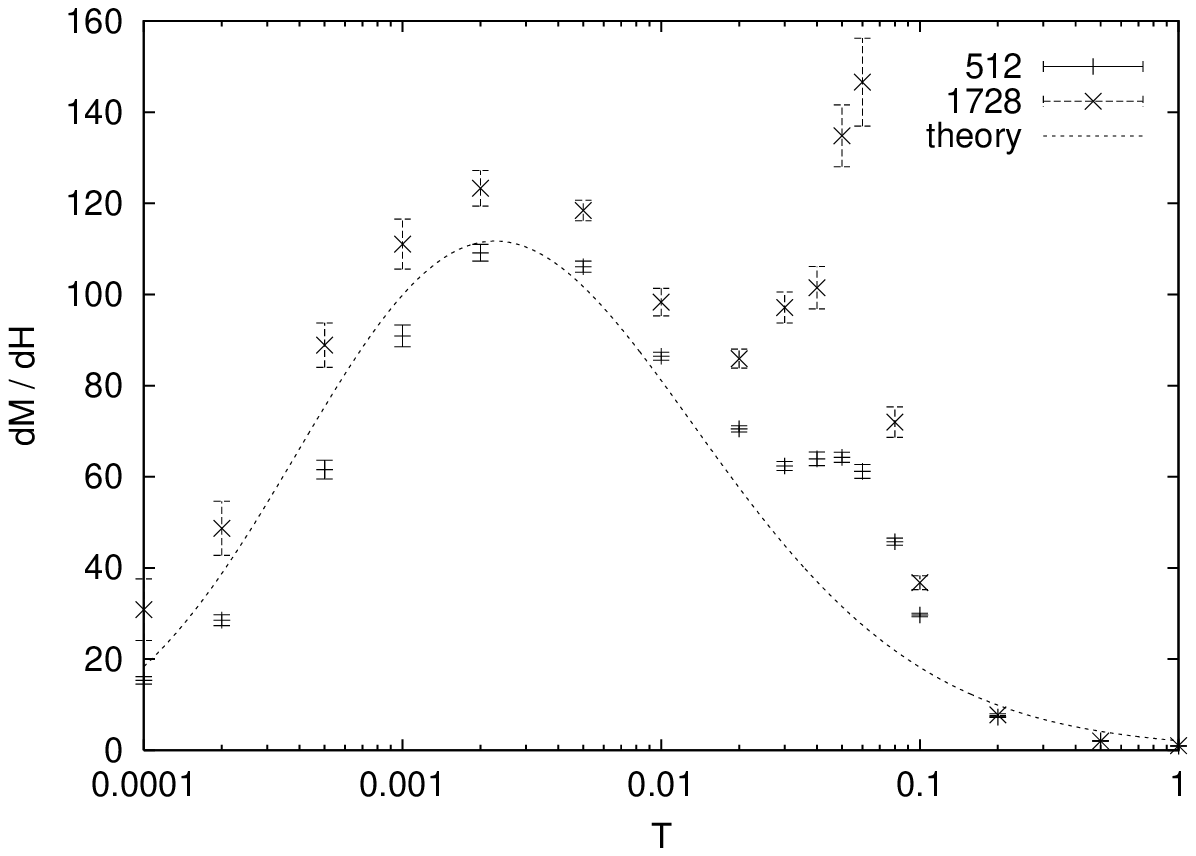,width=3.6in}
}
\caption{ 
\label{nearMIT}
Plot of susceptibility versus temperature $T$ for systems of
various number of Mn spins near the metal-insulator transition
concentration.  
Solid curve is the estimate of the contributions from Mn spins
outside the percolating polaron cluster. } 
\end{figure}

The simulations reported above were for a low concentration of magnetic
ions and a low
concentration of dopants with a ratio such that the interaction between
Mn spins
and between carrier spins can be neglected, and further, the dominant
carrier-Mn interaction can be well estimated through the isolated
carrier wavefunction. However,
as a consequence, the transition temperature is very low which makes
relevant experimental observations almost impossible. 
We can raise the temperature range of interest by increasing the
concentration of dopants ($n_d$); however, working within a
pure spin model restricts the carrier density to lie below the
metal-insulator
transition (MIT) density so the system stays insulating. 
Increasing both the Mn concentration and dopant concentration by a
factor of 2 gets us close to the MIT ($n_d ^ {1/3} a_B = 0.25$). 
Figure~\ref{nearMIT} shows the susceptibility for system of 512 and 1728
Mn spins as a function of $T$.
Both $T_c$ and $T_p$ increase significantly, still leaving
roughly one order of magnitude difference between the two temperatures.
There is a decrease in the height of the low temperature peak, 
because increasing the dopant concentration reduces the number of weakly
coupled Mn spins. 
Our theoretical estimate again fits these results very well, which
implies that the main contribution of the low temperature peak comes
from such isolated Mn spins.

While these simulations were done for the insulating regime, we expect
that
the onset of metallicity, especially just above the MIT, will not change
qualitatively the unusual aspects of the magnetic behavior of doped DMS.
In fact, unusual magnetization curves have been reported in experiments
on metallic $Ga_{1-x}Mn_xAs$, a III-V DMS, where Mn contributes both a
localized
magnetic moment and a carrier (but the system appears to be heavily
compensated, so the carrier density is only about 10\% of the Mn density)
~\cite{Beschoten}. These effects, arising from disorder in the
Mn and dopant positions, have been left out in
recent models~\cite{Konig,Dietl} that have appeared in the literature
since the completion of this work. Preliminary results of a numerical
study
of metallic doped III-V DMS~\cite{Berciu} taking into account the
disorder in the Mn positions shows that qualitative effects seen
in this study remain important at the densities well beyond the MIT.
 
In summary, we have demonstrated via a Monte Carlo study, a
ferromagnetic transition in doped, diluted magnetic
semiconductors in the low doping (insulating) regime.
Though the model studied has no frustration, it has a very
high degree of disorder, characterized by a wide distribution of
exchange couplings. As a result, the system exhibits very unusual
properties, such as thermodynamic properties (e.g., magnetization,
magnetic susceptibility, specific heat) which are very different
from homogeneous systems. In particular, standard expansions used for
homogeneous
systems - critical fluctuations around the transition temperature,
or spin wave expansions around the ordered state ($T=0$), do very poorly
in describing the thermodynamics over most of the phase diagram. There
is considerable thermodynamic entropy down to very low temperatures, 
leading to an unusual peak in the magnetic susceptibility
and specific heat at a temperature well below the transition
temperature.
Our simulations demonstrate the
necessity of incorporating the discrete nature of the quantum 
Heisenberg spins
(despite the ferromagnetic nature of the ordered state),
and the possibility of unusual thermodynamics in the $T \rightarrow 0$
limit also. Some of the qualitative features are expected to survive
in the case of metallic doped DMS as well.

This research was supported by NSF DMR-9809483.

\end{document}